\documentclass[aps,prl,twocolumn,showpacs,10pt,superscriptaddress,preprintnumbers,nofootinbib]{revtex4-1}
\usepackage{epsfig,amssymb,amsmath,psfrag,epstopdf,color}
\pdfoutput=1
\usepackage{graphicx}
\usepackage{subcaption}
\usepackage{ragged2e}
\DeclareCaptionJustification{justified}{\justifying}
 \usepackage[normalem]{ulem}
\usepackage{paralist}
\usepackage{hyperref}
\usepackage[size=tiny]{todonotes}
\allowdisplaybreaks

\def\beq{\begin{equation}}
\def\eeq{\end{equation}}
\def\bsp#1\esp{\begin{split}#1\end{split}}
\newcommand{\be}{\begin{equation}}
\newcommand{\ee}{\end{equation}}
\newcommand{\bea}{\begin{eqnarray}}
\newcommand{\eea}{\end{eqnarray}}

\begin{document}
\preprint{MIT-CTP/5764}

\title{Minimizing Selection Bias in Inclusive Jets in Heavy-Ion Collisions with Energy Correlators --- arXiv note}

\author{Carlota Andres}
\affiliation{Center for Theoretical Physics, Massachusetts Institute of Technology, Cambridge, MA 02139, USA}
\affiliation{Laborat\'orio de Instrumenta\c{c}\"ao e F\'isica Experimental de Part\'iculas (LIP), Av.~Prof.~Gama Pinto, 2, 1649-003 Lisbon, Portugal}

\author{Jack Holguin}
\affiliation{Consortium for Fundamental Physics, School of Physics \& Astronomy, University of Manchester, Manchester M13 9PL, United Kingdom}


\begin{abstract}

This note serves as a companion to a \textit{Letter} \cite{Andres:2024hdd}, where we introduce a new energy correlator-based observable designed to minimize the impact of selection bias due to energy loss in inclusive jets in heavy-ion collisions. Here, we apply the method outlined in the \textit{Letter} to the first-ever measurement of energy correlators in heavy-ion collisions, recently released by the CMS Collaboration \cite{talkEEC,CMS-PAS-HIN-23-004}. 
\end{abstract}

\maketitle
\onecolumngrid

This brief note applies the method introduced in \cite{Andres:2024hdd} to the CMS measurement of the two-point energy correlator (E2C) in inclusive jets in heavy-ion (A-A) collisions  \cite{talkEEC,CMS-PAS-HIN-23-004}. In \cite{Andres:2024hdd}, we introduced a new correlator-based observable, defined as
\begin{align}
    {\rm E2C}/C_{2} \equiv \frac{f^{ \rm AA}_{\rm E2C}(R_{L})}{C_{2}(R_{L})}\,,
    \label{eq:unbiased_spec}
\end{align}
where $f^{\rm AA}_{\rm E2C}(R_L)$ denotes the E2C distribution \cite{Chen:2020vvp,Komiske:2022enw} in heavy-ion jets, with $R_L$ being the angle between detectors, and $C_2(R_L)$ is the so-called \textit{unbiasing function}, designed to correct for leading-order energy loss effects the E2C distribution. For the derivation and explicit functional form of the unbiasing function, we refer the reader to \cite{Andres:2024hdd}.  Once the E2C distribution is measured, the unbiasing function $C_2$  can be directly obtained.  We  note that in proton-proton (p-p) collisions, $C_2$ is unity by construction, making $\text{E2C}/C_2$ identical to E2C for p-p jets.

The CMS measurement uses data in lead-lead (Pb- Pb) collisions  at  center-of-mass energy $\sqrt{{\rm s_{NN}}}=5.02~$TeV.  The E2C distribution was measured for charged particles on inclusive jets with transverse momenta between $120~$GeV and $200~$GeV reconstructed using the anti-$k_T$ algorithm with $R=0.4$. Once a jet is reconstructed, the winner takes all axis was found and a $R=0.4$ cone was drawn around it. All the particles measured inside that cone were correlated. A vacuum reference was analogously found using data on p-p collisions at the same center of mass energy \cite{CMS-PAS-HIN-23-004}. In the following, we will compare the E2C and E2C/$C_2$  across different reconstructed jet transverse momenta and centrality classes. It is important to note that the error bands on the E2C/$C_2$ curves we present are likely overestimated. Since the unbiasing function is derived directly from the E2C distribution, its uncertainties are expected to be highly correlated with those of the E2C itself. However, our current analysis does not account for these correlations in the error propagation. A comprehensive experimental analysis will undoubtedly refine and improve upon the results presented in this \emph{Note}.

\begin{figure}[ht]
    \includegraphics[scale=0.47]{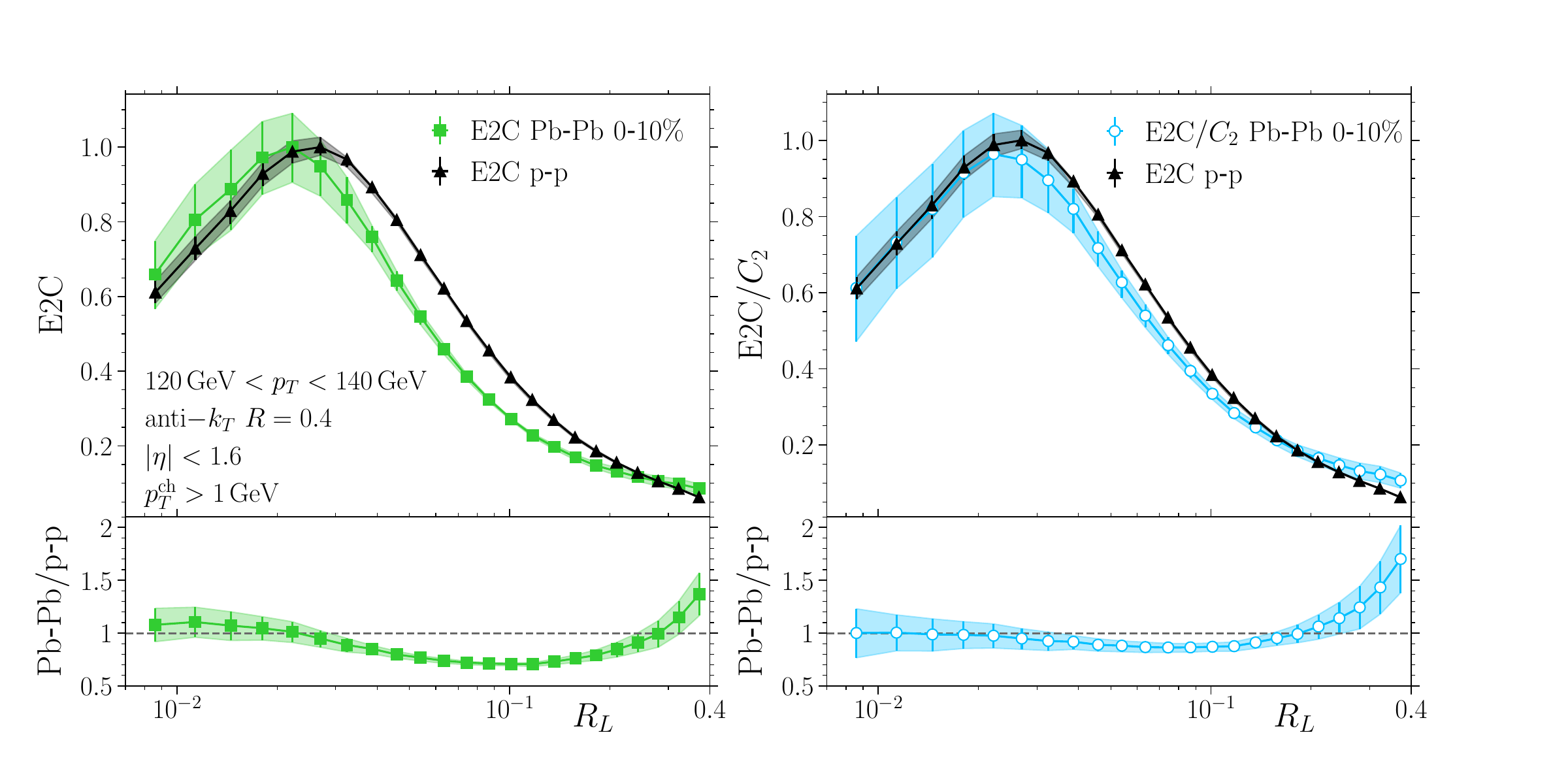}
     \caption{{\bf Left panel}: The top plot shows the E2C for inclusive anti-$k_T$ jets with reconstructed  $p_T$ in the range $120 < p_T < 140$ GeV within the 0-10$\%$ centrality class in $\sqrt{{\rm s_{NN}}}=5.02~$TeV Pb-Pb collisions (green), compared to that of inclusive jets in $\sqrt{{\rm s}}=5.02~$TeV collisions (black) in the same $p_T$ range. The bottom plot shows the ratio of the E2C for Pb-Pb jets to that of p-p jets. {\bf Right panel}: The top plot shows the E2C/$C_2$ for inclusive jets with reconstructed  $p_T$ in the range $120 < p_T < 140$ GeV within the 0-10$\%$ centrality class in $\sqrt{s_{\rm NN}}=5.02~$TeV Pb-Pb collisions (blue), compared to the E2C of inclusive jets in $\sqrt{{\rm s}}=5.02~$TeV collisions (black) in the same $p_T$ range. The bottom plot shows the ratio of the E2C/$C_2$ for Pb-Pb jets to the E2C of p-p jets. We note that $\mathrm{E2C}/C_2\equiv {\rm E2C}$  for p-p jets.}
    \label{fig:fig1}
\end{figure}

In Fig.~\ref{fig:fig1}, we present  a comparison of the E2C (left panel) and E2C/$C_2$ (right panel) for inclusive jets with reconstructed $p_T$ in the range $120 < p_T< 140~$GeV. Results are shown for both p-p and the 0-10$\%$ centrality class of Pb-Pb collisions. In the top left panel, we observe that the Pb-Pb E2C (green) is shifted toward smaller $R_L$ values compared to the p-p E2C (black). This shift results in the overall suppression of the Pb-Pb spectrum relative to the p-p one for $0.03 \lesssim R_L \lesssim 0.2$, as illustrated in the bottom left panel. When the unbiasing function $C_2$ is applied (right panel), the suppression due to energy loss is significantly reduced. Moreover, the ratio between the Pb-Pb E2C/$C_2$  and the p-p one (right bottom panel) becomes completely flat in the hadronization region ($R_L \lesssim 0.03$). This shows that, within the current uncertainties, the observed modifications in the E2C in Pb-Pb compared to p-p in the deep hadronization region (left panel, $R_L<0.015$) are fully attributable to selection bias effects. In the large angular region ($R_L \gtrsim 0.2$), we observe that the enhancement with respect to p-p is larger for the E2C/$C_2$ distribution than for the E2C, indicating that the application of the unbiasing function has effectively mitigated energy loss effects in this region. However, a quantitative description of the remaining enhancement in the E2C/$C_2$ spectrum requires further investigation, as several phenomena---such as medium-induced radiation and medium response ---are predicted to contribute to this behavior \cite{Andres:2022ovj,Andres:2023xwr,Andres:2024ksi,Bossi:2024qho,Yang:2023dwc}. 

\begin{figure}[ht]
    \includegraphics[scale=0.47]{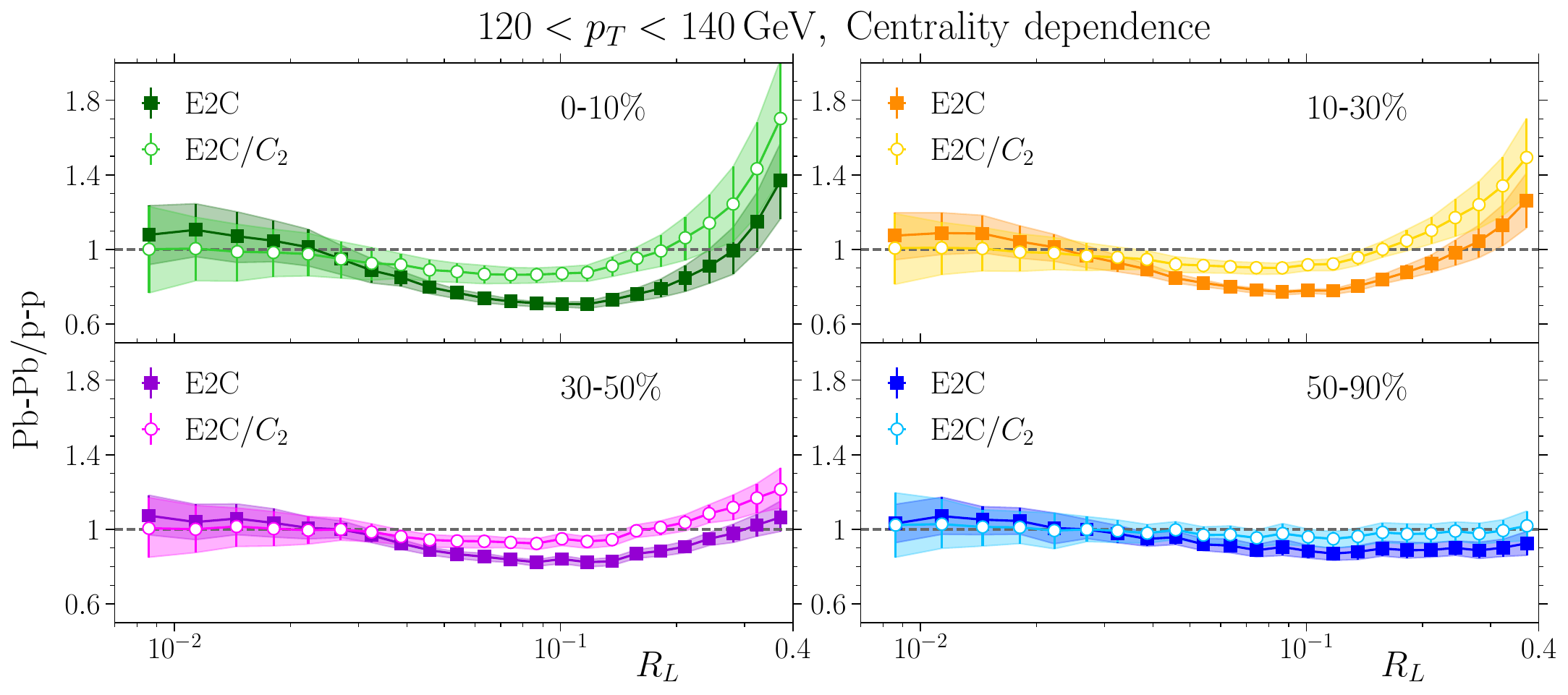}
    \caption{{\bf Top left:} E2C (full squares) and E2C/$C_2$ (empty circles) for inclusive anti-$k_T$ jets with reconstructed  $p_T$ in the range $120 < p_T < 140$ GeV within the 0-10$\%$ centrality class in $\sqrt{ {\rm s _{NN}}}=5.02~$TeV Pb-Pb collisions (squares) divided by the E2C of inclusive jets in $\sqrt{{\rm s}}=5.02~$TeV p-p collisions in the same $p_T$ range.
    {\bf Top right:} As in the top left, but for jets in 10-30$\%$ Pb-Pb collisions. {\bf Bottom left:} As in the top left, but for jets in 30-50$\%$ Pb-Pb collisions. {\bf Bottom right:} As in the top left, but for jets in 50-90$\%$ Pb-Pb collisions.}
    \label{fig:cents}
\end{figure}

In Fig.~\ref{fig:cents}, we analyze the centrality dependence of the E2C and E2C/$C_2$ spectra. The filled squares (open circles) represent the E2C (E2C/$C_2$) within inclusive jets with reconstructed $p_T$ in the range $120 < p_T< 140~$GeV  divided by the corresponding p-p baseline. Each panel corresponds to a different centrality class in Pb-Pb collisions. It is evident that as centrality decreases, the amplitude of the wide-angle enhancement in both the E2C and E2C/$C_2$ ratios diminishes. This observation aligns with the general prediction that both selection bias due to energy loss and other medium effects are reduced in more peripheral collisions \cite{Andres:2022ovj,Andres:2023xwr,Andres:2024ksi,Bossi:2024qho,talkEEC}. The $50\%$-$90\%$ centrality class is  particularly noteworthy. Before applying the unbiasing function (filled dark blue squares), the Pb-Pb/p-p ratio within this class appears to display a nontrivial structure. While, after applying the unbiasing function, the E2C/$C_2$ Pb-Pb/p-p ratio (empty light blue circles) becomes completely flat.  This indicates that a small selection bias was the sole cause of deviation from the vacuum baseline and that the unbiasing function has accurately corrected for this effect.

\begin{figure}[ht]
    \includegraphics[scale=0.47]{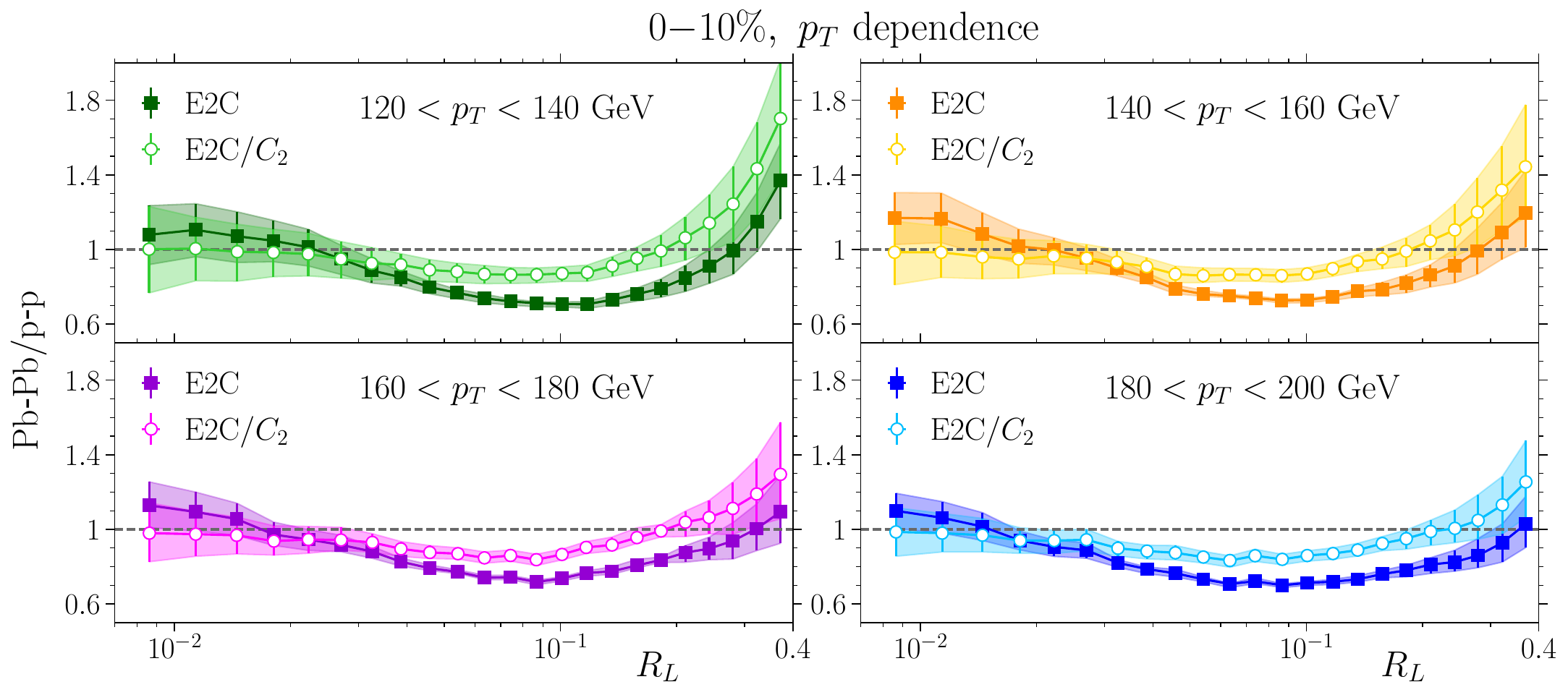}
    \caption{{\bf Top left:} E2C (full squares) and E2C/$C_2$ (empty circles) for inclusive anti-$k_T$ jets with reconstructed $p_T$ in the range $120 < p_T < 140$ GeV within the 0-10$\%$ centrality class in $\sqrt{{\rm s_{NN}}}=5.02~$TeV Pb-Pb collisions (squares) divided by the E2C of inclusive jets in $\sqrt{{\rm s}}=5.02~$TeV p-p collisions in the same $p_T$ range.
    {\bf Top right:} As in the top left, but for jets with $p_T$ in the range 140–160 GeV. {\bf Bottom left:} As in the top left, but for jets with $p_T$ in the range 160–180 GeV. {\bf Bottom right:} As in the top left, but for jets with $p_T$ in the range 180–200 GeV.}
    \label{fig:pts}
\end{figure}

Fig.~\ref{fig:pts} focuses on the $p_T$-dependence of the results. The filled squares (empty circles) represent the E2C (E2C/$C_2$) distribution within inclusive jets in the 0-10$\%$ centrality class in Pb-Pb collisions divided  by the E2C distribution in inclusive  p-p  jets. Each panel corresponds to a different range of reconstructed jet transverse momenta. It can be clearly observed that the wide-angle enhancement decreases in amplitude as the jet $p_T$ increases. Additionally, the wide-angle structure shifts to slightly smaller $R_L$ values as the jet $p_T$ increases. These qualitative behaviors were predicted as consequences of medium-induced radiation prior to the unveiling of the CMS measurement \cite{Andres:2022ovj,Andres:2023xwr,Andres:2024ksi}. However, a comprehensive analysis, including the extraction of scaling laws governing this enhancement, is required to disentangle medium-induced radiation from other effects, such as medium response \cite{Yang:2023dwc,Bossi:2024qho}, and to identify possible signatures of color coherence dynamics \cite{Andres:2022ovj,Andres:2023xwr}.

\begin{figure}[ht]
    \centering
    \includegraphics[scale=0.47]{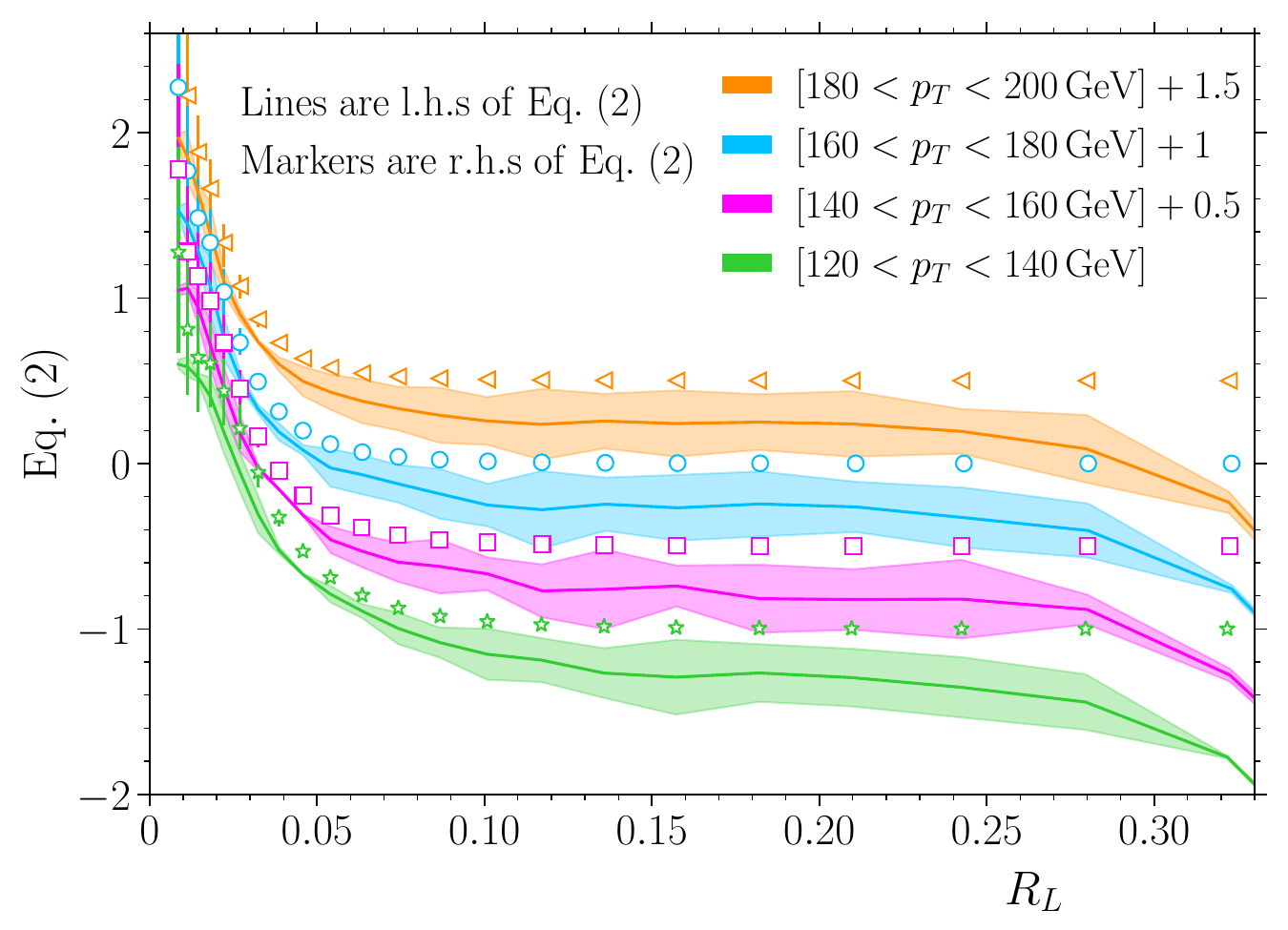}
    \caption{Direct evaluation of Eq.~\eqref{eq:approx} for $p=2$ using the experimental data on the E2C within inclusive anti-$k_T$ jets in $\sqrt{{\rm s}}=5.02$~TeV p-p collisions \cite{CMS-PAS-HIN-23-004}. The left hand side of Eq.~\eqref{eq:approx} is represented by solid curves, while the right-hand side is shown by empty markers. Different colors correspond to results for different reconstructed jet $p_T$ ranges. For clarity, offsets of 0.3, 0.6, and 0.9 have been applied to the results for jets with $p_T$ ranges of $140 < p_T < 160$ GeV, $160 < p_T < 180$ GeV, and  $180 < p_T < 200$ GeV ranges, respectively. Note that, as explained in the main text, error bands are not displayed on the curves representing the l.h.s. of eq.~\eqref{eq:approx}. }
    \label{fig:derivatives}
\end{figure}

Finally, we present in Fig.~\ref{fig:derivatives} the evaluation of the following approximate formula, derived in \cite{Andres:2024hdd},
\begin{align}
    \frac{{\rm d} \ln f^{\rm pp}_{\rm E2C}(R_L) }{ {\rm d} \ln R_{L}} \approx \frac{2}{p+1}\frac{{\rm d} \ln F^{\rm pp}_{\rm E2C} (R_{L},p)}{ {\rm d} \ln R_{L}} - 1\,
    \label{eq:approx}
\end{align}
using the CMS p-p data on the two-point energy correlator within p-p jets \cite{CMS-PAS-HIN-23-004} and for $p=2$. We refer the interested reader to \cite{Andres:2024hdd} for further details on this equation and the role it plays in the derivation of the unbiasing function $C_2$. The derivative on the right hand side of eq.~\eqref{eq:approx} was evaluated analytically, while that on the left hand side was computed numerically. The error bands on the numerical derivatives account for uncertainties from both the numerical method used for their computation and the experimental data. Experimental uncertainties are treated as uncorrelated, as  bin-to-bin correlations from the CMS measurement are not yet available. Notably, the overall uncertainty is dominated by the numerical method. The results shown in Fig.~\ref{fig:derivatives} show good agreement between both sides of eq.~\eqref{eq:approx} over the complete angular range studied, even around the hadronization transition at $0.015<R_L<0.05$.\\

\textit{Conclusions}---This companion  note to the \textit{Letter} \cite{Andres:2024hdd} presents  results on the  correlator-based observable introduced in \cite{Andres:2024hdd}, E2C/$C_2$, using the CMS measurement of the two-point energy (E2C) correlator within inclusive jets in Pb-Pb collisions at the LHC \cite{talkEEC,CMS-PAS-HIN-23-004}. The E2C/$C_2$ observable shows significantly reduced sensitivity to energy loss effects compared to the corresponding  E2C for jets with transverse momenta between 120 GeV and 200 GeV, across different centrality classes in Pb-Pb collisions.  Our current analysis does not account for correlations in the error propagation, and thus we encourage the experiments to perform this measurement.\\

\emph{Acknowledgements.}--- We are especially grateful to Raghav Kunnawalkam Elayavalli and Jussi Viinikainen for helping throughout this project and for granting us the privilege of being the sole authors of this note. We thank Hannah Bossi, Fabio Dominguez, Kyle Lee, Yen-Jie Lee, Cyrille Marquet, Ian Moult,  Krishna Rajagopal, and Carlos A. Salgado for useful discussions. The work of CA was partially supported by the U.S. Department of Energy, Office of Science, Office of Nuclear Physics under grant Contract Number DE-SC0011090 and by OE Portugal, Funda\c{c}\~{a}o para a Ci\^{e}ncia e a Tecnologia (FCT), I.P., project 2024.06117.CERN. CA also acknowledges  financial support by the FCT under contract  2023.07883.CEECIND. The authors would like to express special thanks to the Mainz Institute for Theoretical Physics (MITP) of the Cluster of Excellence PRISMA$^+$ (Project ID 390831469), for its hospitality and support.

\bibliography{refs.bib}{}
\bibliographystyle{apsrev4-1}
\newpage
\onecolumngrid
\newpage


\end{document}